\documentclass[a4paper, oneside, twocolumn,notitlepage, 10pt]{extarticle_ecoc}

\usepackage{ecoc}
\usepackage{acro}

\usepackage{hyperref} 
\usepackage{graphicx}
\graphicspath{../Figure/}
\DeclareGraphicsExtensions{.pdf,.jpeg,.png,.jpg, .svg}
\usepackage[dvipsnames]{xcolor}
\usepackage[font=footnotesize]{caption}
\usepackage{makecell}
\usepackage{psfrag}
\usepackage{tikz}
\usepackage{subcaption}
\usepackage{pgfplots}
\pgfplotsset{compat=1.18} 
\usepgfplotslibrary{groupplots}
\usetikzlibrary{calc}
\usetikzlibrary{patterns}
\newcommand{\gettikzxy}[3]{%
  \tikz@scan@one@point\pgfutil@firstofone#1\relax
  \edef#2{\the\pgf@x}%
  \edef#3{\the\pgf@y}%
}

\usepackage{amsmath}
\usepackage{amsfonts}
\usepackage{mathtools}
\usepackage{amssymb}
\usepackage{mathrsfs}
\usepackage{steinmetz} 
\usepackage{bm}
\usepackage{bbm}
\usepackage{stackengine}
\usepackage{array}
\usepackage{theoremref}
\usepackage{algorithmic}
\usepackage[algo2e]{algorithm2e}
\usepackage{algorithm}

\newlength\myindent
\newlength\myindentt
\usepackage{url}

\DeclareAcronym{AIR}{short=AIR,long= achievable information rate,}

\DeclareAcronym{AWGN}{short=AWGN,long= additive white Gaussian noise,}

\DeclareAcronym{ASE}{short=ASE,long= amplified spontaneous emission,}

\DeclareAcronym{BW-CMA}{short=BW-CMA,long= block-wise CMA,}
\DeclareAcronym{BER}{short=BER,long= bit error rate,}
\DeclareAcronym{BW-DDLMS}{short=BW-DDLMS,long= DDLMS,}
\DeclareAcronym{CMA}{short=CMA,long= constant modulus algorithm,}
\DeclareAcronym{CD}{short=CD,long= chromatic dispersion,}
\DeclareAcronym{CW}{short=CW,long= continuous wave,}
\DeclareAcronym{CPE}{short = CPE,long = carrier phase estimation ,}
\DeclareAcronym{DD}{short=DD,long= decision-directed,}

\DeclareAcronym{DD-Kabsch}{short=DD-{K}absch,long= decision-directed {K}absch,}

\DeclareAcronym{DD-Czegledi}{short=DD-{C}zegledi,long=  decision-directed {C}zegledi,}

\DeclareAcronym{DDLMS}{short=DDLMS,long= decision-directed least mean squares,}

\DeclareAcronym{DP}{short=DP,long= dual-polarization,}

\DeclareAcronym{DP-PDL}{short=DP-PDL,long= dual-polarization PDL,}

\DeclareAcronym{DSP}{short=DSP,long= digital signal processing,}

\DeclareAcronym{DOF}{short=DOF,long= degrees of freedom,}
\DeclareAcronym{EDFA}{short=EDFA,long=  erbium-doped fiber-amplifier,}
\DeclareAcronym{EOFC}{short=EO comb,long=  electro-optic frequency comb,}

\DeclareAcronym{GD}{short=GD,long= gradient descent,}

\DeclareAcronym{GN}{short=GN,long= Gaussian Noise,}

\DeclareAcronym{GMI}{short=GMI,long= generalized mutual information,}

\DeclareAcronym{iid}{short = i.i.d.,long = identically and independently distributed,}
\DeclareAcronym{JCP}{short = JCP,long = joint-channel processing ,}

\DeclareAcronym{LMS}{short=LMS,long= least mean square,}

\DeclareAcronym{LS}{short=LS,long= least-square error,}

\DeclareAcronym{LS-DDLMS}{short= LS-DDLMS,long= LS-DDLMS,}

\DeclareAcronym{LS-SW-ULS}{short= LS-SW-ULS,long = LS-SW-ULS,}

\DeclareAcronym{LO}{short=LO,long =local oscillator,}
\DeclareAcronym{MCMA}{short= MCMA,long= modified {CMA},}

\DeclareAcronym{MI}{short=MI,long= mutual information,}

\DeclareAcronym{MIMO}{short=MIMO,long=  multiple-input multiple-output,}

\DeclareAcronym{ML}{short= ML,long= maximum likelihood}

\DeclareAcronym{MMA}{short=MMA,long= multi-modulus algorithm,}

\DeclareAcronym{MMSE}{short=MMSE,long= minimum mean square error,}

\DeclareAcronym{MSP}{short=MSP,long= master-slave processing,}

\DeclareAcronym{NLI}{short=NLI,long= nonlinear impairments,}

\DeclareAcronym{pdf}{short=pdf,long= probability density function,}

\DeclareAcronym{PDL}{short=PDL,long= polarization-dependent loss,}

\DeclareAcronym{PDM}{short= PDM,long= polarization-division multiplexed,}

\DeclareAcronym{PM}{short= PM,long= polarization multiplexed,}

\DeclareAcronym{PM-16-QAM}{short=PM-$16$-QAM,long= polarization-multiplexed $16$ quadrature amplitude modulation,}

\DeclareAcronym{PMD}{short=PMD,long= polarization-mode dispersion,}

\DeclareAcronym{PN}{short=PN,long= phase noise,}

\DeclareAcronym{PTC}{short=PTC,long= polarization-time code,}

\DeclareAcronym{QAM}{short=QAM,long= quadrature amplitude modulation,}

\DeclareAcronym{QPSK}{short=QPSK,long= quadrature phase-shift keying,}

\DeclareAcronym{RDE}{short=RDE,long= radially directed equalizer,}

\DeclareAcronym{RF}{short=RF,long=radio frequency,}

\DeclareAcronym{RHS}{short=RHS,long=right-hand side,}

\DeclareAcronym{RA}{short= RA, long = reference-assisted}
\DeclareAcronym{SDM}{short=SDM,long=Space-division multiplexing,}

\DeclareAcronym{SNR}{short=SNR,long=signal-to-noise-ratio,}

\DeclareAcronym{SOP}{short=SOP,long= state of polarization,}

\DeclareAcronym{SVD}{short= SVD,long= singular value decomposition}

\DeclareAcronym{SER}{short=SER,long= symbol error rate,}

\DeclareAcronym{SW-Kabsch}{short= SW-Kabsch,long= sliding window Kabsch,}

\DeclareAcronym{SW-LS}{short= SW-LS,long= sliding window least squares,}

\DeclareAcronym{SW-ULS}{short= SW-ULS,long= sliding window unitary least square error,}

\DeclareAcronym{WSS}{short=WSS,long=  wavelength selective switches,}
\DeclareAcronym{WDM}{short=WDM,long=  wavelength-division multiplexing,}
\DeclareAcronym{WD}{short = WD,long = Wrapped Diagonal,}

\DeclareAcronym{WDT}{short = WDD,long = wrapped diagonal distribution,}
\DeclareAcronym{RAT}{short = RAD,long = reference-assisted distribution,}
\DeclareAcronym{EKS}{short = EKS,long = extended {K}alman smoother,}

\definecolor{myMagneta}{rgb}{1, 0, 1}
\renewcommand{\b}[1]{\mathbf{#1}}
\newcommand{\bs}[1]{\boldsymbol{#1}}
\newcommand{\mr}[1]{\mathrm{#1}}

\newcommand{\thetab}{\bs{\theta}}

\newcommand{\thetacb}[1]{\thetab_{#1}^{\mr{c}}}
\newcommand{\thetarb}[1]{\thetab_{#1}^{\mr{r}}}

\newcommand{\Deltab}{\bs{\Delta}}

\DeclareMathAlphabet\mathbfcal{OMS}{cmsy}{b}{n}

\newcommand\norm[1]{\left\lVert#1\right\rVert}
\newcommand\norms[1]{\lVert#1\rVert}

\renewcommand{\vec}[1]{\underline{#1}} 
\newcommand{\rvec}[1]{\underline{\b{#1}}} 

\newcommand\oversetpi[1]{\mathstrut\mkern-1mu#1\mkern-18mu\raise1.25ex%
 \hbox{$\scriptscriptstyle 2\pi$}\mkern3mu}

\newcommand{\Q}{\mr{Q}}

  \newcommand{\T}{\mr{T}}   
  
\DeclareMathOperator*{\argmin}{arg\,min}

\DeclareMathSymbol{\shortminus}{\mathbin}{AMSa}{"39}

\usepackage[mode=tex]{standalone} 
\usepackage{dsfont}
\usepackage{float}
\usepgfplotslibrary{fillbetween}

\newcommand*{\inlineequation}[2][]{%
  \begingroup
    \refstepcounter{equation}%
    \ifx\\#1\\%
    \else
      \label{#1}%
    \fi
    \relpenalty=10000 %
    \binoppenalty=10000 %
    \ensuremath{%
      #2%
    }%
    ~\@eqnnum
  \endgroup
}
\makeatother

\usepackage{paralist}

\begin{document}
\selectlanguage{english}    

\title{Pilot Distributions for Phase Noise Estimation in Electro-Optic Frequency Comb Systems}%


\author{
Mohammad Farsi\textsuperscript{(1)},
  Magnus Karlsson\textsuperscript{(2)}, and
    Erik Agrell\textsuperscript{(1)}
}
\maketitle                  

\begin{strip}
 \begin{author_descr}
 
   \textsuperscript{(1)}
   Dept.~of Electrical Engineering, Chalmers Univ.~of Technology, Sweden, \textcolor{blue}{\uline{farsim@chalmers.se}} \\
   \textsuperscript{(2)}
   Dept.~of Microtechnology and Nanoscience, Chalmers Univ.~of Technology, Sweden\\
 \end{author_descr}
\end{strip}
\renewcommand\footnotemark{}
\renewcommand\footnoterule{}
\setstretch{.98}
\begin{strip}
  \begin{ecoc_abstract}
    We explore the optimal pilot positioning for phase tracking in electro-optic frequency comb setups. We show that, in contrast to previous results for regular multichannel systems, allocating the first and the last channels for pilots is optimal given a fixed pilot overhead. \textcopyright2023 The Author(s)  
  \end{ecoc_abstract}
\end{strip}
\section{Introduction}
Resource sharing among multiple wavelength channels is necessary for co-integrating multiple optical transceivers without exceeding power and heat constraints. The \ac{EOFC}, serving as a multi-wavelength source, provides uniformly spaced carriers for spectral superchannels, enabling co-integration. Moreover, they have applications in metrology, sensing, and telecommunications. 

The \ac{PN} among different comb lines is fully correlated as a result of sharing the same light source. This enables potential resource savings by utilizing joint phase estimation. This becomes more important with high-order modulation formats where \ac{PN} is more detrimental. 

The inherent \ac{PN} correlation between optical comb lines can reduce the computational complexity in digital signal processing through optical comb regeneration techniques \cite{puttnam:2013} or improve the performance of pilot-aided tracking schemes such as \ac{RA} (also known as master--slave) processing \cite{feuer_joint:2012,lundberg_joint:2017} or joint-channel processing \cite{alfredsson_pilot:2018}. 
Previous studies have shown that distributing pilots in time and channel in a grid fashion works best for multichannel systems \cite{alfredsson_pilot:2020}. In \cite{alfredsson_pilot_ecoc:2017,agrell_phase:2018} it was shown that the
optimal pilot placement for space-division multiplexing systems depends on the amount
of \ac{PN} correlation across channels. However, there has been no investigation of pilot placement in the context of \ac{EOFC} systems where the channels are fully correlated. 

In this paper, we explore the use of phase correlation in phase tracking in \acp{EOFC} and analyze pilot placement methods. Optimal pilot positioning is studied and unlike a regular multichannel system, the outer comb lines are found to be optimal for certain estimators. The pilot distribution can substantially impact the bit error rate, with a potential for significant improvement.
\section{System Model}\label{sec:system_model}
Consider uncoded single-polarization transmission in $L$ \ac{WDM} channels generated by an \ac{EOFC}. Also, assume negligible (or already compensated) impairments with the exception of \ac{PN} and amplified spontaneous emission noise. For \acp{EOFC}, most of the \acp{PN} originate from the \ac{CW} laser and the \ac{RF} oscillator, which are shared between all comb lines, resulting in correlated \acp{PN} across the channels \cite{lundberg-freq-comb:2018}. 

The transmitted symbol block in each channel is a random vector of length $N$, where each element is drawn uniformly from a set of equiprobable constellation points. Pilots, with known values and locations, are inserted.
The resulting discrete-time baseband \ac{EOFC} channel model at time $k \in \{1,2,\dots,N\}$ and channel index $m \in\{-M,\dots, M\}$ where $M = (L+1)/2$ and $L$ is the number of channels and odd, is 
\begin{equation}\label{eq:chModel}
    \b{y}_{m,k} = e^{j\thetab_{m,k}}(\b{x}_{m,k}+\b{z}_{m,k}),
\end{equation}
where $\b{y}_{m,k}$, $\b{x}_{m,k}$, $\thetab_{m,k}$, and $\b{z}_{m,k}$ are the received samples, transmitted symbols, total \ac{PN}, and additive, zero-mean, complex, \ac{iid} Gaussian noise, respectively. Here,
\begin{equation}
    \thetab_{m,k} = \thetacb{k} + m\thetarb{k},\label{eq:theta_def}
\end{equation}
where $\thetacb{k}$ and $\thetarb{k}$ denote the \acp{PN} induced by the \ac{CW} laser and \ac{RF} oscillator, respectively. The \ac{PN} sources are statistically independent of each other and all the other random variables and are defined\footnote{The $\mathrm{c/r}$ convention is to avoid redundant repetitions.} as Gaussian random walks \cite{lundberg-freq-comb:2018,ishizawa-phase:2013}
\begin{align}
    \thetab_{k}^{\mathrm{c/r}} &=\thetab_{k-1}^{\mathrm{c/r}}+\Deltab_{k}^{\mathrm{c/r}},
\end{align}
where $\thetab_{0}^{\mathrm{c/r}}$are uniformly distributed in the interval $[-\pi,\pi)$. Moreover, $\Deltab_{k}^{\mathrm{c/r}}$ are independent zero-mean Gaussian variables with variances $\sigma^{2}_\mathrm{c/r} = 2\pi \Delta \nu^\mathrm{c/r} T_{\mathrm{s}}$. These variances describe the drift speed of their corresponding \ac{PN}. Here, $T_{\mathrm{s}}$ is the symbol duration and $\Delta\nu^{\mathrm{c}}$ (typically $1-100$ kHz) and $\Delta\nu^{\mathrm{r}}$ (typically $1-1000$ Hz) denote the \ac{CW} laser and \ac{RF} oscillator linewidths, respectively. 
  
Defining $\rvec{\thetab}_k = (\thetab_{1,k},\dots,\thetab_{M,k})$, we can write the \ac{PN} innovation vector $\rvec{\thetab}_k= \mathrm{T}\cdot [\thetacb{k},\thetarb{k}]^T$, where $\mathrm{T}$ is a $2\times L$ mixing matrix 
\begin{align}
    \mathrm{T} = \begin{bmatrix} 
    1& -M \\
    \vdots& \vdots\\
    1& M \\
    \end{bmatrix}.
\end{align}
\begin{figure}[t]
    \centering
    \includegraphics[height=50pt,width=.46\textwidth]{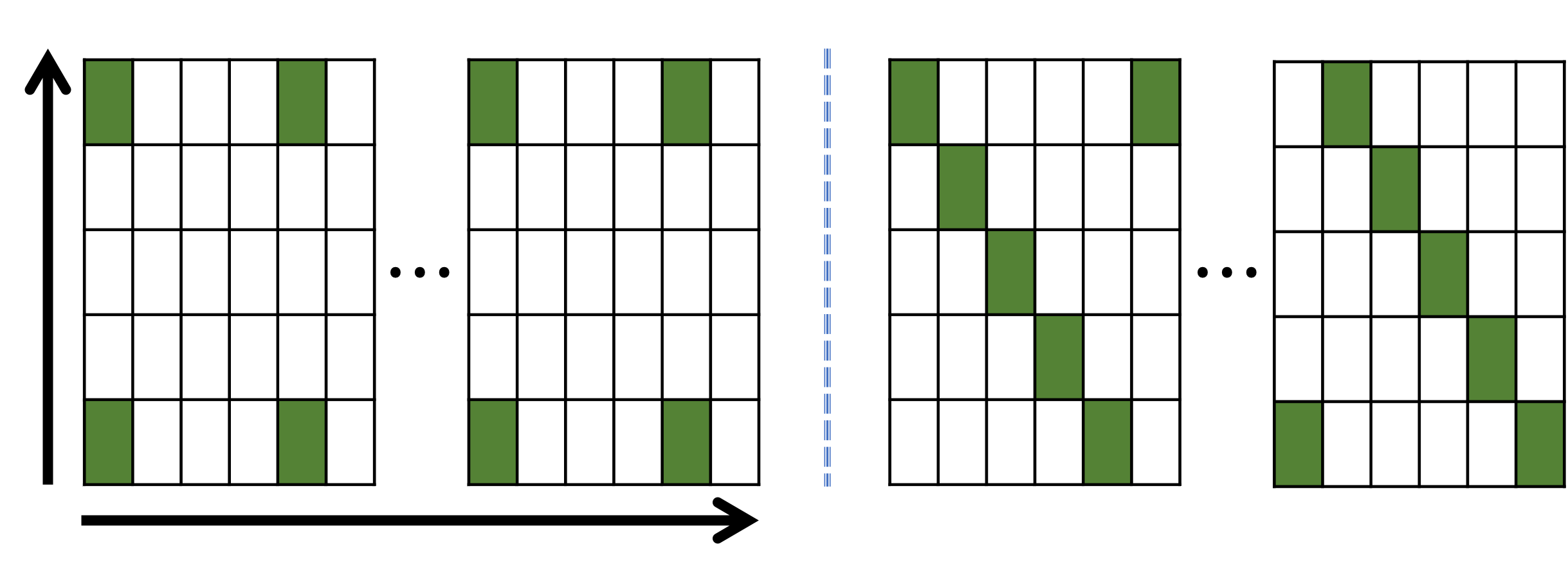}
         \put(-160pt,-6pt){{\small{Time}}}
         \put(-220pt,10pt){\rotatebox{90}{\small{Channel}}}
    \caption{Pilot distributions for $L=5$ ($M=2$) comb lines with $\alpha_{\mr{p}} = 1/5$: RAD (left) and WDD (right), with dark and white blocks representing pilots and data symbols, respectively.}
    \label{fig:three pilots}
\end{figure}
\vspace{-1.1cm}
\section{Reference-Assisted Phase Tracking Scheme}\label{sec:}
The \ac{RA} strategy designates $2\le D\le L$ channels as \textit{reference} channels, where phase tracking is performed and estimated phases are used to correct the \ac{PN} of all the remaining channels, which are designated as \textit{assisted}. 

Let $\mathcal{D} = \{d_1,\cdots,d_D\}$ be a set with distinct elements denoting the reference channel indices. Clearly, all the members of $\mathcal{D}$ are integers in the range $[-M,M]$. The $D\times 1$ \ac{PN} vector of the reference channels can be written as $ \vec{\thetab}_k^{\mathcal{D}} = \Q\cdot[\thetacb{k}, \thetarb{k}]^T$, where 
\begin{align}
    \Q = \begin{bmatrix} 
    1& d_1\\
    \vdots& \vdots\\
    1& d_D\\
    \end{bmatrix}.
\end{align}
Note that we can write $\vec{\thetab}_k = \T\Q^{\dagger}\vec{\thetab}_{k}^{\mathcal{D}}$, where $\Q^{\dagger} =(\Q^T\Q)^{-1}\Q^T$ is the Moore--Penrose inverse of $\Q$.

Without loss of generality, for the reference channels, we define the $D\times 1$ estimated phase vector $\hat{\vec{\thetab}}^{\mathcal{D}}_k$ as
\begin{align}
   \hat{\vec{\thetab}}^{\mathcal{D}}_k =  \vec{\thetab}^{\mathcal{D}}_k+\rvec{w}^{\mathcal{D}}_k , 
\end{align}
where $\rvec{w}^{\mathcal{D}}_k$ is the $D\times 1$ estimation error vector. Its distribution depends on the utilized phase estimation algorithm which is left free to choose. To apply the estimated phases to the assisted channels, we need to form the $L\times 1$ vector of estimated phases as $\hat{\vec{\thetab}}_k = \T{\Q^{\dagger}} \hat{\vec{\thetab}}_k^{\mathcal{D}}$, which gives
\begin{align}
  \hat{\vec{\thetab}}_k = \vec{\thetab}_{k}+\T\Q^{\dagger}\rvec{w}^{\mathcal{D}}_k. \label{eq:est_innovation}
\end{align}
\section{Pilot Placement for \ac{RA} Phase Tracking}
The optimal reference channel index set $\mathcal{D}^{*}$ is the solution to

\begin{align}
   \mathcal{D}^{*}= \argmin_{\mathcal{D}} \mathbb{E}\left[\norm{\rvec{\thetab}_k - \hat{\rvec{\thetab}}_k}^2\right],\label{eq:opt_problem}
\end{align}
where $\hat{\rvec{\thetab}}_k$ is the $L\times 1$ estimated innovation vector in \eqref{eq:est_innovation}, which is obtained from the reference channels. Substituting \eqref{eq:est_innovation} into \eqref{eq:opt_problem}, we obtain
\begin{align}
\mathcal{D}^{*}=\argmin_{\mathcal{D}} \mathbb{E}\left[\norm{\T\Q^{\dagger}\rvec{w}^{\mathcal{D}}_k}^2\right].\label{eq:opt_chan}
\end{align}
Note that the solution of \eqref{eq:opt_chan} is applicable to any estimation algorithm used on the reference channels. For instance,  if an unbiased estimator is selected such that the elements of $\rvec{w}_k^{\mathcal{D}}$ become \ac{iid} and independent of $\mathcal{D}$
, the optimal solution becomes independent of the estimator and can be computed as $\mathcal{D}^{*} =   \argmin_{\mathcal{D}} \norm{\T\Q^{\dagger}}^2_\mathrm{F}$,
where $\norm{\cdot}_\mathrm{F}^2$ denotes the Frobenius norm. 
It can be shown that the optimal set can be formulated as

\begin{align}\label{eq:opt_D}
\mathcal{D}^{*} &= \big\{-M,-M+1,\dots,-M+\lfloor(D+1)/2\rfloor,\nonumber\\
&\qquad M+1-\lfloor D/2\rfloor,\dots,M-1,M\big\}.
\end{align}
\begin{figure*}[h]
\centering
\begin{subfigure}[h]{0.4\textwidth}
 \centering
%
%
\definecolor{mycolor1}{rgb}{0.00000,0.44700,0.74100}%
\newcommand\w{6.5}
\newcommand\h{2.4}
\newcommand\LineWidth{1}
\begin{tikzpicture}

\begin{axis}[%
width=\w cm,
height=\h cm,
at={(0 cm,0cm)},
scale only axis,
xmin=-0.2,
xmax=22.2,
xtick={1,2,3,4,5,6,7,8,9,10,11,12,13,14,15,16,17,18,19,20,21},
xticklabels={{\textcolor{white}{000} $\shortminus$3, ~3},{$\shortminus$3, ~2},{$\shortminus$2, ~3},{$\shortminus$2, ~2},{$\shortminus$3, ~1},{$\shortminus$1, ~3},{$\shortminus$1, ~2},{$\shortminus$2, ~1},{$\shortminus$3, ~0},{~0, ~3},{$\shortminus$1, ~1},{~0, ~2},{$\shortminus$2, ~0},{~1, ~3},{$\shortminus$3,$\shortminus$1},{$\shortminus1$, ~0},{~0, ~1},{~1, ~2},{$\shortminus$2,$\shortminus$1},{$\shortminus$3,$\shortminus$2},{~2, ~3}},
xticklabel style={rotate=90},
ymin=0,
ymax=3.5e-4,
ylabel style={font=\color{white!15!black}},
ylabel={ \small{$\mathcal{E}^{\mathcal{D}}$}}, 
axis background/.style={fill=white},
title style={font=\bfseries, align=center},
xmajorgrids,
xminorgrids,
ymajorgrids,
yminorgrids,
grid style={dotted,white!70!black},
legend style={at={(0.05,0.88)},legend cell align=left,font=\tiny, anchor=north west, align=left, draw=white!15!black}
]
\pgfplotsset{every tick label/.append style={font=\tiny},}

\addplot[ybar, bar width=0.8, fill=red,pattern={crosshatch},pattern color = red , draw=black, area legend] table[row sep=crcr] {%
1	0.000111802048232764\\
};\addlegendentry{$\mathcal{D}^{*}$= \{$\shortminus$3, 3\}}
\addplot[ybar, bar width=0.8, fill=green, draw=black, area legend] table[row sep=crcr] {%
2	0.00011479483183634\\
3	0.000115081063993194\\
};
\addplot[ybar, bar width=0.8,pattern=north east lines,pattern color = teal, draw=black, area legend] table[row sep=crcr] {%
4	0.000118230607430698\\
};
\addplot[ybar, bar width=0.8, fill=blue, draw=black, area legend] table[row sep=crcr] {%
5	0.00012181049760813\\
6	0.00012251216209854\\
};
\addplot[ybar, bar width=0.8, fill=cyan, draw=black, area legend] table[row sep=crcr] {%
7	0.000126108320923926\\
8	0.000126184732255236\\
};
\addplot[ybar, bar width=0.8, fill=Apricot, draw=black, area legend] table[row sep=crcr] {%
9	0.000135754835874784\\
10	0.000136537602895532\\
};

\addplot[ybar, bar width=0.8,pattern=north west lines,pattern color = black, draw=black, area legend] table[row sep=crcr] {%
11	0.000137760882002503\\
};
\addplot[ybar, bar width=0.8, fill=Aquamarine, draw=black, area legend] table[row sep=crcr] {%
12	0.0001455187849807\\
13	0.000145864315340056\\
};
\addplot[ybar, bar width=0.8, fill=DarkOrchid, draw=black, area legend] table[row sep=crcr] {%
14	0.000170653769706147\\
15	0.000172676529035657\\
};
\addplot[ybar, bar width=0.8, fill=brown, draw=black, area legend] table[row sep=crcr] {%
16	0.000178947049518056\\
17	0.000181456450993911\\
};
\addplot[ybar, bar width=0.8, fill=teal, draw=black, area legend] table[row sep=crcr] {%
18	0.000217679006506936\\
19	0.000221664229818473\\
};
\addplot[ybar, bar width=0.8, fill=lightgray, draw=black, area legend] table[row sep=crcr] {%
20	0.000279714009779041\\
21	0.000286957462866202\\
};

\end{axis}
\end{tikzpicture}%
\end{subfigure}
\hfill
\begin{subfigure}[h]{0.55\textwidth}
 \centering
%
%
\definecolor{mycolor1}{rgb}{0.00000,0.44700,0.74100}%
\newcommand\w{6.5}
\newcommand\h{2.5}
\newcommand\LineWidth{1}
\newcommand\BarWidth{.9}
\begin{tikzpicture}

\begin{axis}[%
width=\w cm,
height=\h cm,
at={(0 cm,0cm)},
scale only axis,
xmin=-0.2,
xmax=36.2,
xtick={1,2,3,4,5,6,7,8,9,10,11,12,13,14,15,16,17,18,19,20,21,22,23,24,25,26,27,28,29,30,31,32,33,34,35},
xticklabels={{$\shortminus$3,$\shortminus$2, ~3},{~3, ~2,$\shortminus$3},{$\shortminus$3,$\shortminus$1,  ~3},{3, ~1,$\shortminus$3},{$\shortminus$3, ~0,  ~3},{$\shortminus$3,  ~1,  ~2},{$\shortminus$2,$\shortminus$1, ~3},{$\shortminus$3, ~0, ~2},{$\shortminus$2, ~0, ~3},{$\shortminus$3,$\shortminus$1,  ~2},{$\shortminus$2, ~1, ~3},{$\shortminus$2,  ~2,  ~3},{$\shortminus$3,  $\shortminus$2, ~2},{$\shortminus$2,  $\shortminus$1, ~2},{2, ~1,$\shortminus$2},{$\shortminus$3, ~0, ~1},{$\shortminus$1, ~0, ~3},{$\shortminus$2, ~0, ~2},{$\shortminus$3,$\shortminus$1, ~1},{$\shortminus$1, ~1, ~3},{$\shortminus$1, ~2, ~3},{$\shortminus$3,$\shortminus$2, ~1},{$\shortminus$2, ~0, ~1},{$\shortminus$1, ~0, ~2},{$\shortminus$2,$\shortminus$1, ~1},{$\shortminus$-1, ~1, ~2},{$\shortminus$3,$\shortminus$1, ~0},{0, ~1, ~3},{0, ~2, ~3},{$\shortminus$3,$\shortminus$2, ~0},{$\shortminus$1, ~0, ~1},{0,  ~1, ~2},{$\shortminus$2,$\shortminus$1, ~0},{1,  ~2,  ~3},{$\shortminus$3,$\shortminus$2,$\shortminus$1}},
xticklabel style={rotate=90},
ymin=0,
ymax=3.5e-4,
ylabel style={font=\color{white!15!black}},
ylabel={ \small{$\mathcal{E}^{\mathcal{D}}$}}, 
axis background/.style={fill=white},
title style={font=\bfseries, align=center},
xmajorgrids,
xminorgrids,
ymajorgrids,
yminorgrids,
grid style={dotted,white!70!black},
legend style={at={(0.05,.88)},legend cell align=left,font=\tiny, anchor=north west, align=left, draw=white!15!black}
]
\pgfplotsset{every tick label/.append style={font=\tiny},}

\addplot[ybar, bar width=\BarWidth, fill=red, draw=black, area legend] table[row sep=crcr] {%
1	0.000164322883291079\\
2	0.000164504688879523\\
};\addlegendentry{$\mathcal{D}^{*}$ = \{$\shortminus$3, $\shortminus$2, ~3\}\\$\mathcal{D}^{*}$ = \{$\shortminus$3, ~\,\,2, ~3\}}
\addplot[ybar, bar width=\BarWidth, fill=green, draw=black, area legend] table[row sep=crcr] {%
3	0.000165640906920033\\
4	0.000165926456966682\\
};
\addplot[ybar, bar width=\BarWidth,pattern=north east lines,pattern color = teal, draw=black, area legend] table[row sep=crcr] {%
5	0.000166296116276321\\
};
\addplot[ybar, bar width=\BarWidth, fill=blue, draw=black, area legend] table[row sep=crcr] {%
6	0.000167644884050952\\
7	0.000167937848039011\\
};
\addplot[ybar, bar width=\BarWidth, fill=cyan, draw=black, area legend] 
table[row sep=crcr] {%
8	0.000169022316444881\\
9	0.000169753485452597\\
};
\addplot[ybar, bar width=\BarWidth, fill=Apricot, draw=black, area legend] 
table[row sep=crcr]{%
10	0.000171595327037513\\
11	0.000171904218791721\\
};
\addplot[ybar, bar width=\BarWidth, fill=Aquamarine, draw=black, area legend] table[row sep=crcr] {%
12	0.000172050153289215\\
13	0.000172761322922855\\
};
\addplot[ybar, bar width=\BarWidth, fill=DarkOrchid, draw=black, area legend] table[row sep=crcr] {%
14	0.000175315430144664\\
15	0.000175459369789579\\
};
\addplot[ybar, bar width=\BarWidth, fill=brown, draw=black, area legend] table[row sep=crcr] {%
16	0.000175821783086748\\
17	0.000176861849993283\\
};
\addplot[ybar, bar width=\BarWidth,pattern=north west lines,pattern color = Sepia, draw=black, area legend]
table[row sep=crcr] {%
18	0.000177650160936427\\
};
\addplot[ybar, bar width=\BarWidth, fill=teal, draw=black, area legend] table[row sep=crcr] {%
19	0.000182365621610334\\
20	0.000183369212904957\\
};
\addplot[ybar, bar width=\BarWidth, fill=SpringGreen, draw=black, area legend] table[row sep=crcr] {%
21	0.000185123564344869\\
22	0.000186554634267802\\
};
\addplot[ybar, bar width=\BarWidth, fill=Thistle, draw=black, area legend] table[row sep=crcr] {%
23	0.00018671152735695\\
24	0.000187431194850973\\
};
\addplot[ybar, bar width=\BarWidth,fill=TealBlue,  draw=black, draw=black, area legend] table[row sep=crcr] {%
25	0.000190058473867961\\
26	0.000190103884932092\\
};
\addplot[ybar, bar width=\BarWidth, fill=BrickRed, draw=black, area legend] table[row sep=crcr] {%
27	0.000202188276105901\\
28	0.000202751151294429\\
};
\addplot[ybar, bar width=\BarWidth, fill=Yellow, draw=black, area legend] table[row sep=crcr] {%
29	0.000208908970946528\\
30	0.000210336390932605\\
};
\addplot[ybar, bar width=\BarWidth,pattern=crosshatch dots,pattern color = Sepia, draw=black, area legend] table[row sep=crcr] {%
31	0.000211890124372412\\
};
\addplot[ybar, bar width=\BarWidth, fill=ForestGreen, draw=black, area legend] table[row sep=crcr] {%
32	0.000222779063820077\\
33	0.000224867860054226\\
};
\addplot[ybar, bar width=\BarWidth, fill=lightgray, draw=black, area legend] table[row sep=crcr] {%
34	0.000267556775583661\\
35	0.000267895027555368\\
};
\end{axis}
\end{tikzpicture}%
\end{subfigure}
  \caption{ The mean estimation error $\mathcal{E}^{\mathcal{D}}$ for all possible selections of $D=2$ (left) and $D=3$ (right) reference channels when $L=7$ ($M=3$) and $\alpha_{\mr{p}} = 1/7$. The horizontal axis shows the selected reference channels \scriptsize{$\mathcal{D}$}.}
    \label{fig:RA_channel_choose}
\end{figure*}
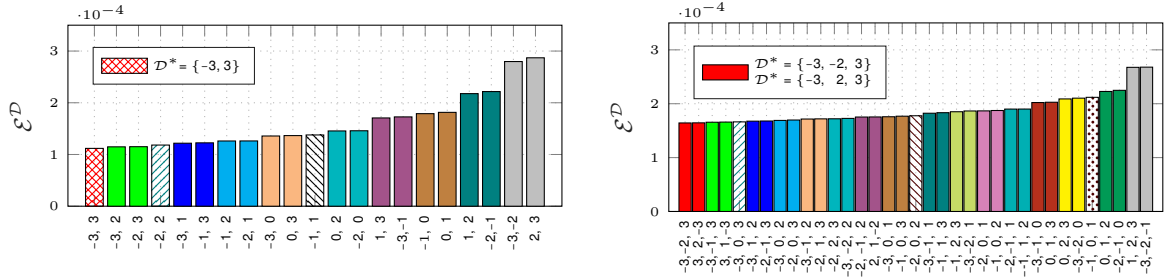
\section{Pilot Distributions}\label{sec:pilot_distribution}
The estimation technique can be done using either an \ac{RA} or a joint-channel processing scheme, depending on the pilot distribution. Let $\alpha_{\mr{p}} \in [0,1]$ represent the average rate across all channels. 

Two pilot distributions for joint-channel processing were examined: \ac{RAT}, in which pilots are positioned at the same time index across a group of reference channels identified by $\mathcal{D}$, and \ac{WDT}, which places pilots along the diagonal in a wrapped pattern. In \ac{RAT}, pilots are only located in the channels specified by $\mathcal{D}$, while in \ac{WDT}, pilots are distributed across all channels using the wrapped diagonal approach. The pilot rate $\alpha_{\mr{p}}$ determines the time spacing between the pilots for both distribution types. Fig.~\ref{fig:three pilots} depicts examples of the distributions. Moreover, the pilot rate of \ac{RAT} is restricted to $\alpha_\mr{p}\le D/L$ and for \ac{WDT} it is limited to $\alpha_\mr{p}\le 1/L$.

\section{Simulation Setup}\label{sec:Results}
Monte Carlo simulations of uncoded $64$-ary quadrature-amplitude modulation (QAM) at a symbol rate of $R_\mathrm{s} = 20$ Gbaud, using system model \eqref{eq:chModel}, were performed with $N=10^5$ symbols per channel. The number of channels $L$ was varied between $11$ to $101$, with a fixed $\alpha_{\mr{p}}=1\%$ pilot overhead and a signal-to-noise-ratio yielding a \ac{BER} of $10^{-3}$ for \ac{PN}-free transmission. The \ac{CW} laser linewidth was fixed at $\Delta\nu^{\mr{c}} = 100\,\text{kHz}$ while the \ac{RF} oscillator linewidth was varied within the practical range of $\Delta\nu^{\mr{r}} \in [1,10^{4}]\,\text{Hz}$. The extended Kalman smoother algorithm from \cite{alfredsson_iterative:2019} was used for phase tracking due to its adaptability to various pilot distributions. For any set of reference channels $\mathcal{D}$, we define the mean estimation error $\mathcal{E}^{\mathcal{D}} =\mathbb{E}[\norms{\T\Q^{\dagger}\rvec{w}^{\mathcal{D}}_k}^2]$. 

Fig.~\ref{fig:RA_channel_choose} illustrates the mean estimation error $\mathcal{E}^{\mathcal{D}}$ for all possible choices of reference channels for $D=2$ (left) and $D=3$ (right) when $L=7$ ($M=3$) and $\alpha_{\mr{p}} = 1/7$. The color-coded bars in the figure indicate channel symmetry, with bars of the same color having identical estimation errors and uniquely hashed bars having no identical twin bars. Small variations in estimation errors are due to randomness. In both cases, the optimal reference channel set indicated by hashed red (left) and solid red (right) aligns with \eqref{eq:opt_D}. Interestingly, for $D=3$, the best channel set is not the uniform choice of channels (first, middle, and last), but rather $\mathcal{D}^{*} = \{-3,-2,3\}$ or its mirror image $\mathcal{D}^{*} = \{-3,2,3\}$. The optimal set of two reference channels $\mathcal{D}^{*} = \{-3,3\}$ results in slightly better $\mathcal{E}^{\mathcal{D}}$ than the case with three reference channels $\mathcal{D}^{*} = \{-3,-2,3\}$. 
\begin{figure}[htb]
    \centering
%
%

%
\definecolor{EKS_wd}{rgb}{0.85000,0.32500,0.09800}%
\definecolor{RAT}{rgb}{0.85000,0.32500,0.09800}%
\definecolor{MS}{rgb}{0.00000,0.44700,0.74100}%
\definecolor{WDT}{rgb}{0.49400,0.18400,0.55600}%
\definecolor{EKS_dist}{rgb}{0.63500,0.07800,0.18400}%
\definecolor{EKS_dp_subopt}{rgb}{0.92900,0.69400,0.12500}%
\definecolor{UBE_dp_opt}{rgb}{0.30100,0.74500,0.93300}%
\definecolor{UBE_dp_subopt}{rgb}{0.49400,0.18400,0.55600}%
\definecolor{awgn}{rgb}{0.92900,0,0.92}%
\newcommand\w{6.2}
\newcommand\h{3.8}
\newcommand\LineWidth{1}
\begin{tikzpicture}

\begin{axis}[%
width=\w cm,
height=\h cm,
at={(0in,0in)},
scale only axis,
xmin=10,
xmax=100,
xlabel style={font=\color{white!15!black}},
xlabel={\scriptsize{ Number of Channels $L$}},
ymode=log,
ymin=0.0008,
ymax=0.03,
yminorticks=true,
xtick={20,40,60,80,100},
xticklabels={{$\text{20}$},{$\text{40}$},{$\text{60}$},{$\text{80}$},{$\text{100}$}},
ylabel style={font=\color{white!15!black}},
ylabel={\scriptsize{BER}},
ytick={0.001,0.01,0.1},
yticklabels={{$\text{10}^{\text{-3}}$},{$\text{10}^{\text{-2}}$},{$\text{10}^{\text{-1}}$}},
axis background/.style={fill=white},
title style={font=\bfseries},
xmajorgrids,
xminorgrids,
ymajorgrids,
yminorgrids,
grid style={dotted,white!70!black},
legend style={at={(0.0,1)}, anchor=north west,font=\scriptsize,legend cell align=left, align=left},
]
\pgfplotsset{every tick label/.append style={font=\scriptsize},}


\addplot [color=MS,line width=\LineWidth pt, mark=triangle, mark options={solid, MS}]
  table[row sep=crcr]{%
11	0.00169727896653522\\
21	0.00144829016576803\\
31	0.00131716555590162\\
41	0.00131723061207243\\
51	0.00128072686419582\\
61	0.00127871183524639\\
71	0.00125206665176193\\
81	0.00123761928697218\\
91	0.00124460750327402\\
101	0.00123118712409116\\
};
\addlegendentry{\ac{RAT} $D =2$}


\addplot [color=MS,line width=\LineWidth pt, mark=o, mark options={solid, ForestGreen}]
  table[row sep=crcr]{%
11	0.0022\\
21	0.0022\\
31	0.0022\\
41	0.0024\\
51	0.0024\\
61	0.0024\\
71	0.0024\\
81	0.0025\\
91	0.0027\\
101	0.0031\\
};
\addlegendentry{\ac{RAT} $D = (L+1)/2 $}
\addplot [color=MS,line width=\LineWidth pt, mark=asterisk, mark options={solid, Red}]
  table[row sep=crcr]{%
11	0.0033\\
21	0.0033\\
31	0.00335\\
41	0.0034\\
51	0.0035\\
61	0.0039\\
71	0.0048\\
81	0.0050\\
91	0.0062\\
101	0.0064\\
};
\addlegendentry{\ac{RAT} $D =L$}
\addplot [color=WDT,line width=\LineWidth pt,densely dashed, mark=square, mark options={solid, WDT}]
  table[row sep=crcr]{%
11	0.00163869074434337\\
21	0.00153953274297429\\
31	0.0014764873241861\\
41	0.00144731319229646\\
51	0.0015686329109784\\
61	0.00154618750993603\\
71	0.00185722260519029\\
81	0.00262048158279981\\
91	0.00648837218011124\\
101	0.129292015522554\\
};
\addlegendentry{\ac{WDT}}

\end{axis}

\begin{axis}[%
width=\w cm,
height=\h cm,
at={(0in,0in)},
scale only axis,
xmin=0,
xmax=1,
ymin=0,
ymax=1,
axis line style={draw=none},
ticks=none,
axis x line*=bottom,
axis y line*=left
]
\end{axis}
\end{tikzpicture}%
    \caption{\ac{BER} of various pilot distribution for a fixed pilot rate $\alpha_{\mr{p}} = 1\%$ versus the number of channels $L$. The \ac{RF} oscillator linewidth is $\Delta\nu^{\mr{r}} = 100\text{Hz}$.}
    \label{fig:pilot_dist_ber}
\end{figure}
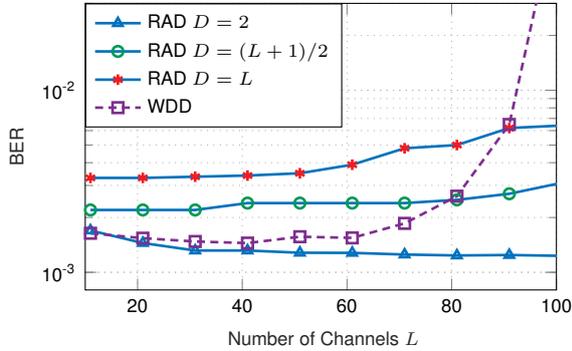

Fig.~\ref{fig:pilot_dist_ber} shows the \ac{BER} as a function of $L$ for different pilot distributions at a fixed pilot rate of $\alpha_{\mr{p}} = 1\%$ and \ac{RF} oscillator linewidth of $\Delta\nu^{\mr{r}} = 100\,\text{Hz}$. For each $D$ and $L$, the reference channels are optimally selected according to \eqref{eq:opt_D}. Our findings suggest that the best \ac{RAT} is achieved when $D=2$ reference channels are used (i.e., $\mathcal{D}=\{-M,M\}$), while the poorest performance is observed when pilots are placed in all channels (i.e., $\mathcal{D} = \{-M,-M+1,\dots,M\}$). Additionally, it is observed that increasing the number of reference channels $D$ leads to higher \ac{BER}, suggesting that $D=2$ is optimal at a fixed pilot rate. For $L>40$, the \ac{PN} of the \ac{RF} oscillator becomes more significant at the outer channels, leading to faster phase changes. This makes it harder to accurately estimate the phase in the outer channels using the inner channels, which may explain why \ac{WDT} performs poorly at higher values of $L$. 

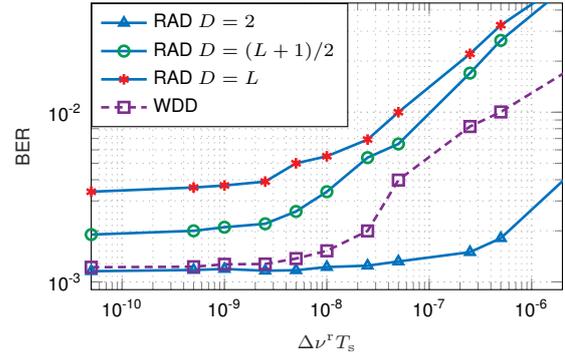
\begin{figure}[htb]
    \centering
%
%
\definecolor{EKS_wd}{rgb}{0.85000,0.32500,0.09800}%
\definecolor{RAT}{rgb}{0.85000,0.32500,0.09800}%
\definecolor{MS}{rgb}{0.00000,0.44700,0.74100}%
\definecolor{WDT}{rgb}{0.49400,0.18400,0.55600}%
\definecolor{EKS_dist}{rgb}{0.63500,0.07800,0.18400}%
\definecolor{EKS_dp_subopt}{rgb}{0.92900,0.69400,0.12500}%
\definecolor{UBE_dp_opt}{rgb}{0.30100,0.74500,0.93300}%
\definecolor{UBE_dp_subopt}{rgb}{0.49400,0.18400,0.55600}%
\definecolor{awgn}{rgb}{0.92900,0,0.92}%
\newcommand\w{6.2}
\newcommand\h{3.8}
\newcommand\LineWidth{1}
\definecolor{mycolor1}{rgb}{1.00000,0.00000,1.00000}%
\definecolor{mycolor2}{rgb}{0.00000,1.00000,1.00000}%
\begin{tikzpicture}

\begin{axis}[%
width=\w cm,
height=\h cm,
at={(0in,0in)},
scale only axis,
xmode=log,
xmin=5e-11,
xmax=2e-06,
xminorticks=true,
xlabel style={font=\color{white!15!black}},
xlabel={\scriptsize{$\Delta\nu^{\mathrm{r}}T_\mathrm{s}$}},
ymode=log,
ymin=0.0009,
ymax=0.044,
yminorticks=true,
ylabel style={font=\color{white!15!black}},
ylabel={\scriptsize{BER}},
ytick={0.001,0.01,0.1},
yticklabels={{$\text{10}^{\text{-3}}$},{$\text{10}^{\text{-2}}$},{$\text{10}^{\text{-1}}$}},
xtick={1e-11,1e-10,1e-9,1e-8,1e-7,1e-6,1e-5},
xticklabels={{$\text{10}^{\text{-11}}$},{$\text{10}^{\text{-10}}$},{$\text{10}^{\text{-9}}$},{$\text{10}^{\text{-8}}$},{$\text{10}^{\text{-7}}$},{$\text{10}^{\text{-6}}$},{$\text{10}^{\text{-5}}$}},
xtick style={font=\scriptsize},
axis background/.style={fill=white},
title style={font=\bfseries},
xmajorgrids,
xminorgrids,
ymajorgrids,
yminorgrids,
grid style={dotted,white!70!black},
legend style={at={(0.0,1)}, anchor=north west,font=\scriptsize,legend cell align=left, align=left},
]


\pgfplotsset{every tick label/.append style={font=\scriptsize},}

\addplot [color=MS,line width=\LineWidth pt, mark=triangle, mark options={solid, MS}]
  table[row sep=crcr]{%
5e-11	0.00115600920726133\\
5e-10	0.00117539506334073\\
1e-09	0.00119137989203779\\
2.5e-09	0.00116383157024074\\
5e-09	0.00117131383048191\\
1e-08	0.00122300944669366\\
2.5e-08	0.00124613643289365\\
5e-08	0.00131925852161421\\
2.5e-07	0.00149747235644944\\
5e-07	0.00180934656741108\\
2.5e-06	0.00443936104218362\\
5e-06	0.00807063797832049\\
};
\addlegendentry{\ac{RAT} $D =2$}

\addplot [color=MS,line width=\LineWidth pt, mark=o, mark options={solid, ForestGreen}]
table[row sep=crcr]{%
5e-11	0.0019\\
5e-10	0.0020\\
1e-09	0.0021\\
2.5e-09	0.0022\\
5e-09	 0.0026\\
1e-08	 0.0034\\
2.5e-08	 0.0054\\
5e-08	0.0065\\
2.5e-07	0.0170\\
5e-07	0.0265\\
2.5e-06	0.0587\\
5e-06	0.0726\\
};
\addlegendentry{\ac{RAT} $D =(L+1)/2$}
\addplot [color=MS,line width=\LineWidth pt, mark=asterisk, mark options={solid, Red}]
  table[row sep=crcr]{%
5e-11	0.0034\\
5e-10	0.0036\\
1e-09	0.0037\\
2.5e-09	 0.0039\\
5e-09	0.0050\\
1e-08	0.0055\\
2.5e-08	0.0069\\
5e-08	 0.0100\\
2.5e-07	0.0220785714285714\\
5e-07	0.032591156462585\\
2.5e-06	0.0627969387755102\\
5e-06	0.0926085034013605\\
};
\addlegendentry{\ac{RAT} $D =L$}
  
\addplot [color=WDT,line width=\LineWidth pt,densely dashed, mark=square, mark options={solid, WDT}]
  table[row sep=crcr]{%
5e-11	0.00121994603260762\\
5e-10	0.00122810848166884\\
1e-09	0.00126960093106336\\
2.5e-09	0.00127470246172662\\
5e-09	0.00137163154432856\\
1e-08	0.00152399726013793\\
2.5e-08	0.00199707920364426\\
5e-08	0.00398123452960826\\
2.5e-07	0.00823727151094414\\
5e-07	0.010035050916677\\
2.5e-06	0.0182311700802709\\
5e-06	0.0185651502876923\\
};
\addlegendentry{\ac{WDT}}


\end{axis}
\end{tikzpicture}%
    \caption{BER of different pilot distribution types is shown for a fixed pilot rate $\alpha_{\mr{p}}=1\%$ and $\Delta\nu^{\mr{c}} = 100\,\text{kHz}$.} 
    \label{fig:ber_vs_deltar}
\end{figure}
Fig.~\ref{fig:ber_vs_deltar} depicts the \ac{BER} versus normalized \ac{RF} oscillator linewidth $\Delta\nu^{\mr{r}} T_{\mr{s}}$ for $L=51$ comb lines using $\alpha_{\mr{p}} = 1\%$. Again, for each $D$, the set of reference channels are chosen according to \eqref{eq:opt_D}. For $\Delta\nu^{\mr{r}} T_{\mr{s}}\le 10^{-8}$, \ac{WDT} and \ac{RAT} with $D=2$ perform similarly, but \ac{RAT} with $D=2$ outperforms the other pilot distributions at higher values of $\Delta\nu^{\mr{r}} T_{\mr{s}}$. 
Furthermore, it is evident that the optimal performance is achieved when $D=2$ reference channels are utilized, as it outperforms the other values across the range of studied $\Delta\nu^{\mr{r}} T_\mr{s}$. 

Our hypothesis is that when there are time slots available to fill with pilots (i.e., $\alpha_\mr{p}\le 2/L$), the optimal number of reference channels is 2 ($D=2$). Following the same logic, we speculate that $D_{opt} = \max \{ 2, \lceil \alpha_p L \rceil \}$.

\section{Conclusion}
Several types of pilot distributions were compared
through Monte Carlo simulations in terms of the
resulting \ac{BER} performance for tracking correlated
\ac{PN} in \ac{EOFC} systems. It was shown theoretically and confirmed by simulations that for certain phase estimators, the optimal reference channels are the outer channels (first and last), contrary to regular \ac{WDM} channels where pilots are placed in all channels. 

\vspace{.3cm}
\noindent\footnotesize{\textbf{Acknowledgements:} 
This work was supported by the Knut and Alice Wallenberg Foundation, grant No.~2018.0090.
}
\clearpage
\printbibliography
\end{document}